# Kubernetes Misconfigurations in the Wild: Taxonomy, Evolution, and Automated Repair with Large Language Models

Mostafa Anouar Ghorab
Laval university
Quebec, Canada
mostafa-anouar.ghorab.1@ulaval.ca

Ahmad Abdel Latif
University of Calgary
calgary , Canada
ahmad.abdellatif@ucalgary.ca

Mohamed Aymen Saied
Laval university
Quebec, Canada
mohamed-aymen.saied@ift.ulaval.ca

## Abstract

Kubernetes has become a central platform for orchestrating cloud-native applications, yet its declarative configuration model frequently introduces security misconfigurations that threaten system reliability and operational stability. Although automated detection tools are widely available, a systematic understanding of misconfiguration patterns and scalable correction mechanisms remains limited. This paper presents a comprehensive empirical study of Kubernetes security misconfigurations based on 2,662 developer-reported issues from Stack Overflow. From this dataset, we derive a structured taxonomy that captures recurring security weaknesses across configuration object types and misconfiguration categories. Using this taxonomy, we analyze how severity levels vary across objects and categories, and examine how security misconfigurations evolve between incubator and stable project stages. Our findings reveal that while some operational issues decrease as projects mature, critical security misconfigurations often persist or reappear, highlighting enduring risk patterns in cloud-native systems. Building on this empirical foundation, we evaluate the effectiveness of Large Language Models (LLMs) in automatically correcting Kubernetes security misconfigurations under progressively enriched contextual conditions. Results demonstrate that contextual grounding significantly improves correction accuracy, with the best standalone model achieving 89.06%. To further enhance structural correctness and schema compliance, we introduce Kubecurity, a schema-guided validation framework that enforces compliance with official Kubernetes specifications. By combining contextual LLM reasoning with deterministic schema enforcement, the proposed hybrid approach achieves 98.50% correction accuracy while substantially reducing newly introduced misconfigurations. Overall, this work advances both the understanding and automated remediation of Kubernetes security misconfigurations.

## 1 Introduction

Over the past decade, Kubernetes has emerged as the de facto standard for container orchestration in cloud-native environments, enabling automation, scalability, and elasticity across modern distributed systems [6][23][21] [26]. Yet, the very flexibility that empowers these capabilities also introduces substantial configuration complexity [22][36]. As Kubernetes adoption continues to expand, ensuring correct and secure configurations has become a critical challenge.

The consequences of misconfiguration are both tangible and costly. Industry reports document severe financial impacts: one SaaS company incurred over $100,000 in monthly cloud expenses due to an autoscaling misconfiguration that triggered uncontrolled resource consumption [35]. The Cloud Native Computing Foundation estimates that remediating a misconfiguration in production can cost up to $15,900 per workload—over 600 times the cost of addressing the same issue during development ($25) [10]. Even seemingly minor errors, such as misaligned resource quotas or overly permissive access controls, can propagate across clusters and systematically degrade system reliability [31][8].

Despite configuration analysis tools, misconfigurations remain widespread [31][39]. This persistence largely stems from Kubernetes' continuous evolution: each release introduces new APIs, deprecates legacy features, and expands the configuration surface [37]. These changes require administrators to continually adapt clusters and deployment strategies to maintain compatibility, creating a management burden of interdependent parameters where subtle inconsistencies can cascade into critical failures.

Current solutions predominantly employ rule-based detection mechanisms that encode expert knowledge into predefined policies. While effective at identifying known anti-patterns, these tools exhibit limited interoperability—rules and labels are often platform-specific, hindering systematic reasoning across environments. More fundamentally, most approaches emphasize detection over remediation, delegating corrective actions to human operators. In dynamic environments characterized by continuous integration and deployment, this manual, iterative correction process becomes both time-consuming and error-prone.

Addressing these limitations requires automated mechanisms capable of context-aware correction, not merely detection. Recent advances in machine learning, particularly large language models (LLMs), offer new capabilities for reasoning over configuration artifacts and generating specification-compliant repairs. In this work, we present a systematic investigation of Kubernetes security misconfigurations, spanning empirical characterization to automated correction. We address the following research questions:

- **RQ1:** How can empirical evidence from Stack Overflow inform a taxonomy of Kubernetes security misconfigurations?
- **RQ2:** How does security misconfiguration severity vary across object types and categories?
- **RQ3:** How do security misconfigurations evolve between incubator and stable project stages?
- **RQ4:** How effectively can LLMs correct Kubernetes security misconfigurations, and how does **Kubecurity** enhance their reliability?

Together, these questions aim to advance both structured understanding and reliable, automated correction of Kubernetes security misconfigurations.

## 2 Related work

Extensive research has examined Kubernetes misconfiguration detection through empirical studies and specialized tooling. However, while detection mechanisms have advanced considerably, severity assessment and automated remediation remain underexplored.

Prior work analyzed configuration errors across manifests and runtime environments [8][39][1][47][4][25]. Rahman et al. [31] conducted a large-scale study across 2,039 manifests from 92 repositories, identifying 1,051 unique misconfigurations in eleven categories, with 94% of practitioners reporting production incidents caused by such errors. These findings established a foundational understanding of misconfiguration prevalence and operational impact in real-world deployments. Traditional detection relies on rule-based static analysis tools, including Datree [11], KubeScore [1], Snyk[38], and KubeLinter [2], which enforce predefined policies to identify common anti patterns. Although effective at catching known violations, these tools cannot adapt to evolving threats or contextual nuances, often yielding false alarms or missing subtle issues that depend on deployment context.

To overcome these limitations, recent research has explored data-driven and reasoning-based approaches. LLM-powered systems including Malul [24], Cohen [34], and GenKubeSec [25] introduced pipelines capable of detecting and reasoning about misconfigurations with notable precision and recall gains over static baselines. KubeGuard [34] further leveraged LLMs to jointly analyze manifests and runtime logs for privilege escalation detection, illustrating how AI-assisted systems achieve deeper contextual understanding beyond static rule sets. Prior work by the authors also explored LLM-based detection of Kubernetes misconfigurations [15] and ML-driven security analysis in adjacent domains [7][14], confirming the broader applicability of semantic reasoning to configuration security.

Despite these advances, the field lacks a standardized and fine-grained taxonomy of Kubernetes misconfigurations. The SLI-KUBE taxonomy [30] and Rahman et al. [31] provide classification frameworks that link misconfiguration types to operational and security impacts, but both cover only a subset of issues detected by current tooling and were constructed without large-scale practitioner evidence. Severity assessment remains even more neglected: existing tools assign fixed severity levels to individual rules, which fails to reflect actual risk, since violations of the same rule can differ substantially in impact depending on the affected object type and deployment context. Remediation is similarly underserved most detection tools delegate corrective actions entirely to practitioners. Polaris [12] is among the few that validate configurations and suggest fixes through predefined correction rules, but shares the adaptability limitations of all static systems. While emerging AI frameworks show promise for intelligent correction, rigorous empirical validation of their accuracy and compliance with official Kubernetes schemas remains absent. The present work addresses these gaps by introducing a data-driven taxonomy derived from developer discussions, an empirically grounded severity mapping across object types, and an automated correction framework that integrates schema validation with LLM-based semantic reasoning.

## 3 Methodology

This section outlines the methodology for the four research questions, integrating empirical data collection, mixed analyses, and automated correction evaluation.

### 3.1 RQ1 – Constructing a Taxonomy of Kubernetes Misconfigurations

This research question aims to build a comprehensive taxonomy of Kubernetes misconfigurations based on real-world practitioner discussions. Stack Overflow was used as the main data source, offering structured question–answer exchanges that reflect practical configuration issues [19].

*3.1.1 Data Collection* Data collection followed a structured, iterative process led by two practitioners to ensure coverage and relevance. Using the Stack Exchange API, we extracted all Stack Overflow tags, which were jointly reviewed to identify those related to Kubernetes. The refined subset was then used to filter Kubernetes-specific posts while minimizing cross-technology noise. Practitioners also curated configuration and security-related keywords. Using these tags and keywords, we queried Stack Overflow's archive (2015–2024), retaining posts with at least one Kubernetes tag and one keyword, yielding 681 posts (Set A). To address missing or incorrect tags, we additionally collected posts using the identified keywords only. Since keyword-based retrieval may introduce noise, we trained a supervised Support Vector Machine (SVM)[18] classifier using Set A as positive examples and an equal number of unrelated posts as negative examples. To validate the classifier, we adopted a k-fold cross-validation strategy to assess its generalization capability and mitigate the risk of overfitting. The final model achieved an accuracy of 97.8%, with a precision of 97.14%, a recall of 98.55%, and an F1-score of 97.84%, demonstrating strong and well-balanced performance in distinguishing relevant from non-relevant posts. The validated classifier was subsequently applied to filter genuine misconfiguration discussions from unrelated posts, yielding 1,981 curated posts (Set B). This two-phase design ensured both high precision and broader coverage. Sets A and B formed the empirical foundation for taxonomy construction.

*3.1.2 Taxonomy Construction* To identify recurring themes and latent structures, we employed hierarchical BERTopic [17], a transformer-based topic modeling framework leveraging contextual embeddings from BERT to extract both coarse and fine-grained topics. We used the `paraphrase-MiniLM-L3-v2` model from Sentence Transformers for its balance between semantic expressiveness and computational efficiency. To determine the optimal granularity, we tested several `min_topic_size` configurations (2–10) and computed intra- and inter-topic distances to assess clustering quality. Lower intra-topic distances indicate higher semantic cohesion, while higher inter-topic distances reflect better distinctiveness. A minimum topic size of five produced the most coherent and well-separated topics.

We then applied BERTopic's hierarchical reduction procedure via `hierarchical_topics()`, which recursively merges semantically similar clusters based on cosine similarity of their c-TF-IDF representations. This produced a multi-level hierarchy capturing both general themes and specialized subtopics.

The hierarchical structure was manually refined through iterative labeling and consolidation. Two practitioners independently assigned descriptive titles to parent topics and subtopics during the initial phase. Multiple review sessions were held to reconcile naming inconsistencies and resolve overlaps by consensus. This human-in-the-loop process ensured that the resulting taxonomy preserved both data-driven organization and contextual fidelity to Kubernetes configuration practices, following iterative refinement strategies established in prior empirical software studies [5][32][33].

## 3.2 RQ2 – Analyzing Misconfiguration Severity Across Object Types and Categories

While taxonomy-based classification helps characterize Kubernetes misconfigurations, their severity cannot be inferred solely from type. Because each object plays a distinct operational role within the cluster, this study examines how misconfiguration severity varies across object types and taxonomy categories.

*3.2.1 Dataset Construction* We collected 10,000 Kubernetes configuration files from public GitHub repositories spanning 2015–2024. For each year, we queried the GitHub API using a set of keywords composed of the tags and keywords identified in RQ1, and iteratively collected files until reaching 1,000 parsable YAML configurations returned by the API, without applying any popularity or star-based filtering. Only files that were syntactically correct and successfully validated against the Kubernetes schema using Kubeval were retained. Each validated configuration was then analyzed using the three state-of-the-art Kubernetes misconfiguration detection tools identified in prior comparative research [16]. These tools were selected for their high detection accuracy and broad rule coverage, ensuring comprehensive and reliable identification of misconfigurations across diverse Kubernetes objects.

*3.2.2 Mapping and Quantitative Analysis* To ensure conceptual coherence, All detected misconfigurations were mapped to the taxonomy categories defined in RQ1 using a manually constructed correspondence table aligning tool labels with our taxonomy. We then built a two-dimensional matrix crossing Kubernetes object types with taxonomy categories, aggregating both frequency and severity to compare risks across configuration classes.

Severity levels were standardized into three categories (*Low*, *Medium*, *High*) following NIST SP 800-30 [20], OWASP Risk Rating [29], and CVSS v3.1 [13]. Because tools use different native scales (*Datree*: Low–Medium–High–Critical; *Snyk* and *kube-score*: Low–Medium–High), we normalized them by merging Datree's *Critical* into *High*. Finally, a majority rule was applied at each matrix intersection: a level exceeding 50% was marked dominant; otherwise the two most frequent levels were retained.

## 3.3 RQ3 – Evolution of Misconfigurations Across Project Maturity Levels

The third research question investigates the evolution of Kubernetes configurations to understand how misconfigurations emerge, persist, and are resolved as systems mature. It further examines whether configuration issues are effectively mitigated over time and whether new ones continue to appear even in stable versions.

*3.3.1 Data Selection* For this study, we focused on Kubernetes configuration data drawn from the official Helm chart repository. We selected this repository because it provides access to the configuration files of each project in both its incubator and stable phases. This characteristic makes it particularly suitable for analyzing the evolution of Kubernetes configurations as projects transition from early development to production maturity.

To ensure comparability, we included only projects that exist in both the incubator and stable phases and that preserve a consistent architectural structure between these two versions. This design constraint allowed us to isolate the impact of configuration evolution from architectural or structural differences, ensuring that any observed misconfiguration changes are attributable to configuration evolution rather than project redesign.

Following this selection process, 73 projects out of an initial pool of 282 were retained for analysis, representing a total of 664 Kubernetes configuration files. These files encompass a variety of application domains, offering a diverse and representative dataset for assessing how Kubernetes configurations evolve throughout the Helm chart lifecycle.

*3.3.2 Misconfiguration Analysis* To achieve a rigorous and comprehensive assessment of Kubernetes misconfigurations across both incubator and stable Helm charts, this study employed three well-established detection tools widely adopted in the Kubernetes ecosystem: Snyk, Datree, and Kube Score. These tools were selected for their complementary analysis capabilities [16].

To ensure consistency and avoid redundancy, identical misconfigurations reported by multiple tools were merged and counted once. When tools identified different categories in the same file, each unique detection was retained. If the same issue appeared with varying severity levels, the highest severity was kept to reflect a conservative assessment.

By integrating results from multiple detection tools through a unified consolidation process, this study provides a comprehensive view of misconfigurations across the Helm chart lifecycle. The analysis reveals how misconfigurations emerge, persist, or are corrected between incubator and stable phases, highlighting the Kubernetes objects most susceptible to recurrent or unresolved issues.

## 3.4 RQ4 – Evaluating the Capability of LLMs to Correct Misconfigurations

This research question seeks to assess the ability of Large Language Models (LLMs) to rectify Kubernetes misconfigurations through two principal experiments: first, by leveraging various types of contextual information; and second, by employing contextual information augmented through schema-guided correction mechanisms.

*3.4.1 Prompt-Based Experimental Design* To evaluate the corrective capabilities of LLMs, we conducted experiments analyzing how different levels of contextual enrichment affect model performance.

**Misconfiguration-only :** The LLM receives only the faulty YAML file without additional context. This configuration evaluates the model's intrinsic capacity to identify and correct misconfigurations purely from syntactic and structural patterns, representing a zero-shot reasoning scenario.

**Type-augmented :** The input combines the misconfigured file with the taxonomy-defined misconfiguration type (e.g., TLS not enforced or Missing NetworkPolicy ). This minimal semantic guidance directs the model's reasoning toward a specific conceptual domain, improving the precision of repair suggestions.

**Type with detailed Description :** In this configuration, the prompt was further enriched with a detailed natural-language description of the misconfiguration (e.g., "The container lacks a liveness probe, which prevents Kubernetes from detecting and restarting unresponsive containers.").

For each experiment, the model generated a corrected configuration candidate. The resulting outputs were automatically validated against the official Kubernetes JSON schemas to ensure syntactic correctness and structural integrity. They were then reassessed using the detection tools employed in RQ2 and RQ3 namely Datree, Snyk, and KubeScore to verify the absence of any misconfigurations. A correction was considered successful only if it satisfied both schema validation and tool-based verification. In this study, a configuration is considered corrected if it becomes both syntactically valid under Kubernetes JSON schema validation and no longer triggers any misconfiguration detection rules from the employed tools.

*3.4.2 Schema-Guided Rule-Based Correction with Kubecurity* As a complementary component to the LLM-based correction process, we developed Kubecurity, a schema-driven, rule-based correction engine designed to enforce deterministic compliance with official Kubernetes specifications. Each Kubernetes object type is formally defined by a JSON Schema that specifies its structure, required attributes, permitted data types, and contextual constraints. Kubecurity leverages this formal specification to automatically detect and repair configuration inconsistencies. By reasoning over the configuration tree, Kubecurity reconstructs the hierarchy, aligning each element with its parent and ensuring proper field ordering and indentation. Missing required fields are supplemented with default schema values, and type conflicts are reconciled by converting invalid entries into compliant forms. Invalid or obsolete sections are removed, and the configuration is revalidated against the reference schema for full compliance. Beyond correction, Kubecurity acts as a safeguard for LLM-based outputs. While LLMs provide context-aware fixes, their results may violate Kubernetes syntax or structure. Kubecurity runs before and after the LLM stage: initially to repair deterministically resolvable issues, and later to revalidate and fix residual inconsistencies. This dual validation ensures all configurations remain compliant with Kubernetes standards.

# 4 Results

In this section, we present and discuss our findings for each research question.

## 4.1 RQ1 – Taxonomy Presentation

To address the first research question, we implemented a multi-stage process encompassing data collection, topic modeling, and iterative refinement, as described in the methodology section.

The final taxonomy comprises five major categories and twenty-one subcategories, each capturing a distinct dimension of how configuration faults manifest in cloud-native systems. Together, these categories form a structured framework for understanding, diagnosing, and remediating configuration weaknesses across the Kubernetes ecosystem. Figure 1 provides a comprehensive overview of the proposed taxonomy.

Across the 2,613 relevant Stack Overflow posts analyzed, the distribution of misconfigurations was as follows: Access and Privileges (23.8%), Resource Management and Probes (21.5%), Image and Network Security (19.0%), Encryption and Permission Management (20.3%), and Filesystem Configuration (15.4%).

**Access and Privileges:** Emerged as the most frequently discussed category, this category includes misconfigurations related to authorization, privilege escalation, and RBAC policy management. When improperly configured, these issues can enable users or workloads to perform unauthorized operations, resulting in sensitive data exposure or cluster compromise.

**Snippet 2: allowPrivilegeEscalation**

```
apiVersion: v1
kind: Pod
metadata:
  name: privileged-pod
spec:
  containers:
  - name: vulnerable-container
    image: myapp:latest
    securityContext:
      allowPrivilegeEscalation: true  # Should
          be false
```

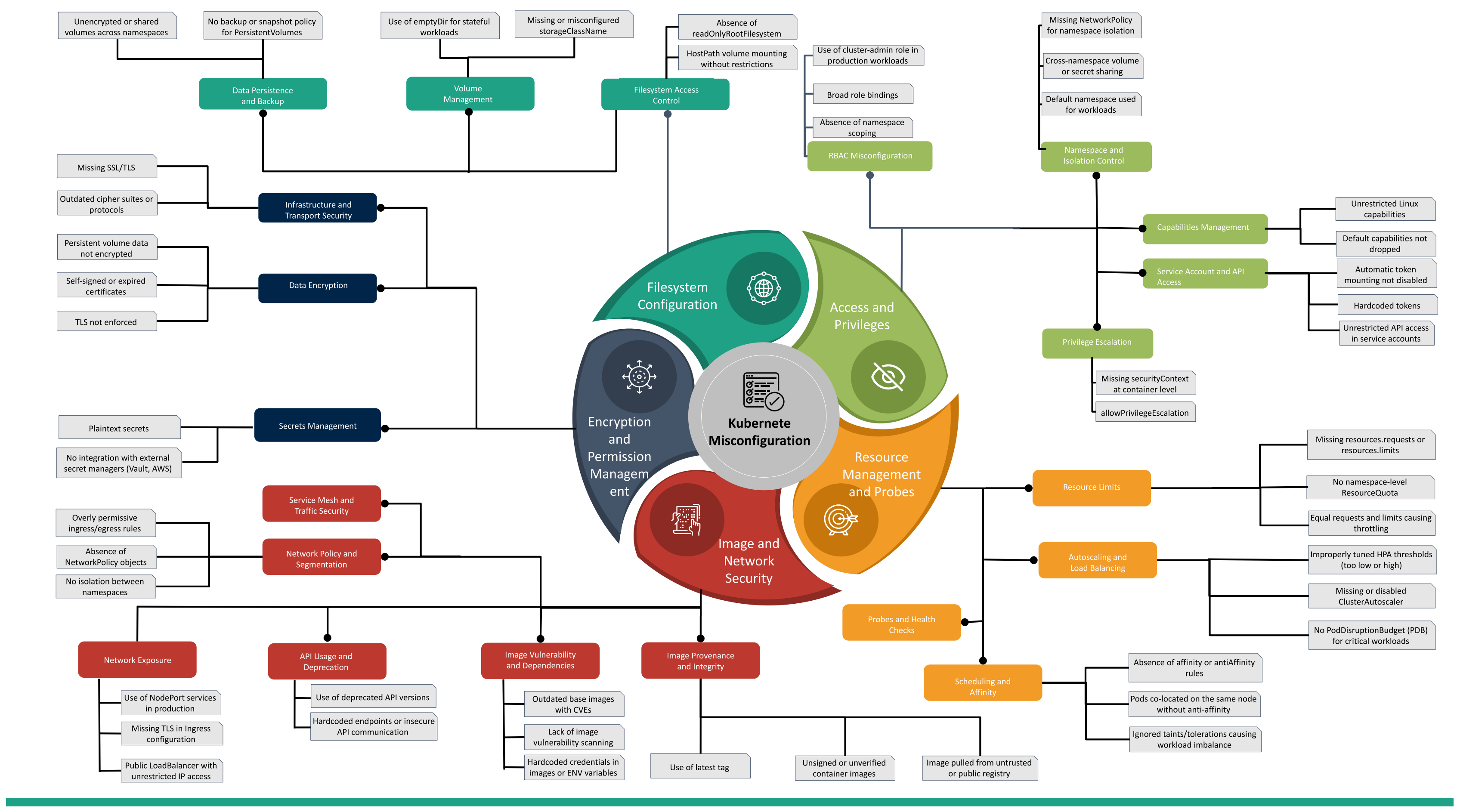

**Figure 1: Kubernetes Misconfiguration Taxonomy**

Here, `allowPrivilegeEscalation: true` enables the container to execute privileged system calls. If exploited, it may grant root access to the underlying host. Following the principle of least privilege, this field should be explicitly set to `false` and combined with restrictive RBAC policies to limit system exposure.

**Resource Management and Probes:** This category covers inefficiencies and risks caused by improper resource allocation, scheduling, or health monitoring configurations. Missing resource requests and limits may destabilize multi-tenant clusters, resulting in workload starvation or node overcommitment.

**Snippet 3: Missing resources.requests or resources.limits**

```
apiVersion: v1
kind: Pod
metadata:
  name: unbounded-pod
spec:
  containers:
  - name: unbounded-container
    image: myapp:stable
    resources: {}  # Missing requests and limits
```

This configuration omits both `resources.requests` and `resources.limits`, granting unrestricted resource consumption. Without these controls, a single container can monopolize CPU or memory, degrading the performance of co-located workloads. Defining appropriate resource boundaries prevents this issue and ensures predictable scheduling.

**Image and Network Security:** This category focuses on vulnerabilities linked to image provenance, network exposure, and security of inter-component communications. Misconfigurations here often facilitate the deployment of unverified images or expose services to public networks without encryption.

**Snippet 4: Use of latest tag**

```
apiVersion: v1
kind: Pod
metadata:
  name: insecure-image-pod
spec:
  containers:
  - name: app
    image: myapp:latest  # Using 'latest' tag
        introduces inconsistency
```

Using the generic `latest` tag leads to unpredictable deployments, as different nodes may pull different image versions. This inconsistency can reintroduce known vulnerabilities or regressions. To ensure stability and traceability, images should be version-pinned and verified through trusted registries.

**Encryption and Permission Management:** This category encompasses misconfigurations in data protection, secrets handling, and transport-layer security. Weak encryption or plaintext secret storage undermines confidentiality and exposes sensitive credentials.

**Snippet 5: Plaintext secrets**

```
apiVersion: v1
kind: Secret
metadata:
  name: plaintext-secret
type: Opaque
data:
  password: myplaintextpassword  # Should be
      encrypted
```

In this example, the secret value is stored as plaintext rather than encoded or encrypted, making it visible to anyone with read access to the manifest. All sensitive data should be base64-encoded or managed through external secret management systems (e.g., HashiCorp Vault, AWS Secrets Manager) to ensure confidentiality and compliance.

**Filesystem Configuration:** Representing the least frequent category, this class encompasses misconfigurations affecting storage and data persistence, including filesystem access control, volume management, and backup policies. Such errors can lead to unauthorized host access, data loss, or insecure persistence across namespaces. Improper filesystem configuration is among the most severe misconfiguration types, as it can directly compromise system integrity.

**Snippet 1: Absence of readOnlyRootFilesystem**

```
apiVersion: v1
kind: Pod
metadata:
  name: insecure-filesystem-pod
spec:
  containers:
  - name: app
    image: nginx:1.25
    securityContext:
      readOnlyRootFilesystem: false  # Should be
          true for safety
```

In this example, setting `readOnlyRootFilesystem: false` allows write access to the container's root filesystem. This increases the attack surface by permitting file modifications and potential privilege escalation through tampered binaries. To mitigate this risk, it is recommended to set `readOnlyRootFilesystem: true`, ensuring a secure, immutable filesystem environment.

## 4.2 RQ2 – Misconfiguration Severity Analysis

This section presents the results of the second research question, examining how the severity of Kubernetes misconfigurations varies across object types and taxonomy categories.

Across the 10,000 validated configuration files collected between 2015 and 2024, the three detection tools jointly reported 23,736 misconfigurations. Because several tools often identified the same issues, duplicates were consolidated to preserve analytical integrity. After deduplication in which each unique misconfiguration was counted once regardless of how many tools detected it,the final set contained 14,375 distinct misconfigurations.

Following the normalization process described in the methodology, all severity levels were harmonized into a three-tier classification (Low, Medium, High). The consolidated results indicate that 60.19% of the misconfigurations were Low, 38.27% were Medium, and 1.54% were High severity. This distribution suggests that most configuration errors affect reliability or maintainability, while nearly a quarter represent severe risks to security or cluster stability. This distribution indicates that the majority misconfigurations have limited operational impact, while a very small subset poses serious risks to cluster security or stability. Despite their low frequency, high-severity misconfigurations remain critical, as they often involve privilege mismanagement, insecure networking, or the exposure of sensitive information. Table 1 summarizes the empirical severity patterns across Kubernetes object types and taxonomy categories.

Building on this distribution, a deeper examination across Kubernetes object types and taxonomy categories reveals several empirical observations and severity patterns.

**Pods and Deployments :** exhibited the highest proportion of Medium to High-severity misconfigurations, predominantly within *Access and Privileges* and *Image and Network Security*. Common patterns included privileged containers, unsafe `hostPath` mounts, and unsecured image pull configurations. Although these High-severity cases account for only a small share of the total corpus, their operational implications ranging from container escape to unverified image execution make them disproportionately impactful.

**Services and Ingresses :** presented a mixed severity profile, dominated by Low and Medium levels within *Image and Network Security*. Typical misconfigurations involved missing TLS termination, open network ports, and unrestricted external exposure. Despite being less severe on average, these configurations increase the external attack surface, demonstrating that even moderate misconfigurations can have compounding security effects when applied to exposed components.

**PersistentVolumes** and **PersistentVolumeClaims :** were largely affected by *FileSystem Configuration* issues, mainly of Low and Medium severity. The majority of errors involved incorrect access modes or misaligned persistence settings. In a limited number of cases ($\approx 2\%$), High-severity faults were identified, typically when host-level paths were insecurely mounted—posing risks of unauthorized data access.

**ServiceAccounts, Roles, and RoleBindings :** formed the most critical cluster of objects in terms of severity distribution. More than half of their misconfigurations were classified as Medium or High, concentrated in *Access and Privileges* and *Encryption and Permission Management*. These reflect deep structural vulnerabilities, such as over-permissive role definitions, missing namespace scoping, or unbounded credential exposure misconfigurations capable of compromising the cluster's security perimeter.

**ConfigMaps and Secrets:** showed a predominance of Low and Medium severities under *Encryption and Permission Management*. Frequent issues included plaintext storage of sensitive credentials and incorrect use of ConfigMaps to hold confidential data. Although High-severity cases were rare, they correspond to direct violations of encryption policies and thus carry significant potential impact.

**Jobs, CronJobs, and HorizontalPodAutoscalers :** were primarily affected by *Resource Management and Probes* issues, almost entirely Low or Medium in severity. These misconfigurations typically involved missing readiness or liveness probes, suboptimal scaling parameters, or loosely defined restart policies. While not directly security-critical, their recurrence indicates systemic weaknesses in reliability-oriented configuration practices.

The two-dimensional analysis reveals that misconfiguration severity in Kubernetes is inherently contextual, relational, and non-uniform. Severity does not arise solely from the intrinsic nature of a misconfiguration, but rather from its interaction with the operational semantics of the affected object, such as responsibilities, privilege boundaries, and exposure level. A privilege escalation that remains benign within a confined namespace may become catastrophic when applied to a cluster-wide role. Likewise, an omitted probe that merely reduces scaling efficiency in a HorizontalPodAutoscaler (HPA) can trigger service unavailability when it affects a Pod. These observations empirically validate the necessity of an intersectional severity analysis, in which risk is characterized through the joint consideration of object type and misconfiguration category, rather than through a flat or category-only assessment.

## 4.3 RQ3 – Misconfiguration Evolution Across Maturity Levels

To examine how Kubernetes misconfigurations evolve with project maturity, we analyzed the transition from incubator to stable Helm chart versions. The results reveal heterogeneous correction patterns across taxonomy categories.

Operational issues received the greatest attention. *Resource Management and Probes* exhibited the highest correction rate (55%), followed by *FileSystem Configuration* (42.86%), indicating that availability and performance concerns are prioritized as projects progress toward stability.

In contrast, security-related categories showed significantly lower remediation rates. *Access and Privileges* achieved a correction rate of only 16.28%, while *Encryption and Permission Management* reached 17.22%. Similarly, *Image and Network Security* recorded a correction rate of 24.14%.

Overall, out of 336 misconfigurations detected in incubator versions, only 72 were corrected in stable releases, corresponding to a global correction rate of 21.43%. Additionally, 21 new misconfigurations (6.25%) emerged in stable versions, demonstrating that configuration evolution is neither linear nor purely corrective.

A closer inspection of unresolved cases reveals that the majority of persistent issues lie within security-critical domains. More than 80% of misconfigurations in *Encryption and Permission Management* and *Access and Privileges* remained uncorrected. These findings suggest that while developers actively prioritize operational stability and performance, security refinements are often deprioritized.

Overall, Kubernetes configuration evolution appears dynamic and non-linear, marked by both progress and regression. The persistence of security weaknesses highlights the limitations of manual remediation and motivates the investigation of automated correction mechanisms based on Large Language Models and schema-guided validation.

[!t]

## 4.4 RQ4 – LLMs Capability in Correcting Misconfigurations

This section investigates the corrective capabilities of five LLMs in repairing Kubernetes misconfigurations. The models Mistral-7B [42], GPT-3.5-Turbo [27], GPT-5.4 [28], LLaMA 3.1-70B [44], DeepSeek R1 [41], and Qwen 2.5-7B [45] were selected to capture diversity across proprietary and open-weight paradigms, as well as across model scales and reasoning strategies.

**Table 1: Empirical Assessment of Kubernetes Misconfigurations Across Object Types and Taxonomy Categories**

| Kubernetes Object Type | FileSystem Configuration | Access & Privileges | Resource Mgmt. & Probes | Image & Network Security | Encryption & Permission Mgmt |
|---|---|---|---|---|---|
| Pod | **Low / Medium** | **Medium / High** | **Medium** | **Medium / High** | **Low** |
| Deployment | **Low / Medium** | **Medium / High** | **Medium** | **Medium / High** | **Low** |
| Service | **Low** | **Medium** | **Medium** | **Medium** | **Low** |
| Ingress | – | **Medium** | **Medium** | **Medium** | **Low** |
| ConfigMap | **Low / Medium** | – | – | **Medium** | **Medium** |
| Secret | – | – | – | – | **Medium / High** |
| PersistentVolumeClaim (PVC) | **Medium** | **Medium** | **Low / Medium** | **Low** | **Low** |
| PersistentVolume (PV) | **Medium** | **Medium** | **Low / Medium** | **Low** | **Low** |
| ServiceAccount | **Medium / High** | **Medium / High** | **Medium** | **Medium** | **Medium / High** |
| Role / ClusterRole | **Medium** | **Medium / High** | **Medium** | **Medium** | **Medium / High** |
| RoleBinding / ClusterRoleBinding | **Medium** | **Medium / High** | **Medium** | **Medium** | **Medium / High** |
| Job | **Low** | – | **Medium** | **Low** | **Low** |
| CronJob | **Low** | – | **Medium** | **Low** | **Low** |
| HorizontalPodAutoscaler (HPA) | **Low** | – | **Medium** | **Low** | **Low** |

**Table 2: Performance Comparison Across Experiments**

| Model | Experiment 01 | | | | Experiment 02 | | | | Experiment 03 | | | | Experiment 04 | | | |
|---|---|---|---|---|---|---|---|---|---|---|---|---|---|---|---|---|
| | Not Parsed | Not Corrected | Corrected | Emerged | Not Parsed | Not Corrected | Corrected | Emerged | Not Parsed | Not Corrected | Corrected | Emerged | Not Parsed | Not Corrected | Corrected | Emerged |
| **Mistral-7B** | 59.24% | 25.32% | 15.71% | 359 | 1.21% | 12.82% | 86.24% | 507 | 1.15% | 12.00% | 87.12% | 359 | 1.13% | 2.10% | 97.04% | 46 |
| **GPT-3.5-Turbo** | **61.01%** | 24.86% | 14.40% | 217 | 1.46% | 11.21% | 87.60% | 306 | 1.32% | 9.89% | 89.06% | 217 | 0.80% | 1.94% | 97.53% | 21 |
| **LLaMA 3.1-70B** | 51.52% | **41.10%** | 7.65% | 310 | **5.04%** | 11.56% | 83.66% | 224 | **4.56%** | 9.56% | 86.14% | 310 | 0.48% | 1.77% | 98.02% | 43 |
| **DeepSeek R1** | 33.27% | 28.62% | 38.38% | **409** | 2.11% | 10.62% | 87.54% | 514 | 2.62% | 8.91% | 88.74% | **409** | 0.32% | 1.45% | 98.50% | 91 |
| **Qwen 2.5-7B** | 53.95% | 31.56% | 14.76% | 286 | 4.47% | **14.44%%** | 81.37% | 308 | 3.76% | **14.27%** | 82.24% | 286 | **1.44%** | **4.53%** | 93.98% | **127** |
| **GPT-5.4** | 30.49% | 25.13% | 44.38% | 391 | 1.83% | 9.27% | **89.90%** | **519** | 1.52% | 7.86% | **90.62%** | 392 | 0.21% | 1.19% | **98.60%** | 79 |

GPT-3.5-Turbo served as a strong baseline due to its consistent performance in structured reasoning and prompt adherence [3]. GPT-5.4 extends the proprietary model family by introducing more advanced reasoning and instruction-following capabilities, enabling a more rigorous assessment of recent improvements in corrective accuracy under complex configuration constraints. LLaMA 3.1-70B, an open-weight model, enables the evaluation of transparent architectures and their ability to leverage contextual cues in configuration repair [46]. Mistral-7B and Qwen 2.5-7B, both lightweight yet high-performing models, provide insights into the trade-off between parameter efficiency and corrective accuracy [43][9]. Finally, DeepSeek R1 introduces a reasoning-oriented design, emphasizing structured problem decomposition and logic-driven inference [40], which aligns well with the declarative nature of Kubernetes configurations.

The evaluation was performed on a corpus of 10,000 Kubernetes configuration files extracted from Helm charts and deployment domains. After normalization and deduplication, 13,495 distinct misconfigurations were identified across five taxonomy categories.

*4.4.1* ***Prompt-Based Correction*** The experiments assessed how progressively richer contextual information influences LLMs' ability to correct misconfigurations. In Experiment 1, models received only the faulty configuration file. Experiment 2 added the misconfiguration type to provide structural guidance. Experiment 3 further included a brief natural-language explanation of the issue, aligning with Kubernetes best practices. This stepwise design enabled a systematic analysis of how increasing context affects reasoning quality, correction accuracy, and the introduction of new misconfigurations. The comparative results are presented in Table 2.

The results summarized in Table 2 show a clear trend: increasing contextual information consistently improves correction accuracy and reasoning stability across all models. In Experiment 1, where only the faulty configuration was provided, performance was limited due to the absence of structural and semantic cues, leading to high parsing failure rates (over 50%) and structurally invalid fixes. DeepSeek R1 achieved the highest correction rate (38.38%), while LLaMA 3.1-70B performed poorly (7.65%), indicating that model size alone does not ensure structural understanding. GPT-5.4 already shows a stronger baseline in this setting, reaching 44.38% corrected outputs, outperforming all other models even under minimal context.

In Experiment 2, adding the misconfiguration type significantly improved results, with all models surpassing 80% correction accuracy and parsing errors nearly disappearing. GPT-3.5-Turbo (87.6%) and DeepSeek R1 (87.54%) achieved comparable performance, while GPT-5.4 further improved to 89.90%, confirming its stronger ability to exploit minimal semantic guidance for stabilizing reasoning and reducing irrelevant or hallucinated corrections.

Experiment 3 further enhanced performance by including a descriptive explanation of the issue and its risks. Correction rates exceeded 86% for all models, with GPT-3.5-Turbo and DeepSeek R1 approaching 89%, while GPT-5.4 reached 90.62%, maintaining a consistent advantage. The richer context enabled more precise, risk-aware fixes and significantly reduced the introduction of new misconfigurations, confirming that descriptive context mitigates overcorrection and unnecessary edits even for already strong models.

#### 4.4.2 *Schema-Guided Rule-Based Correction with Kubecurity*

Experiment 4 assessed the impact of integrating Kubecurity, a schema-guided validation mechanism, into the correction pipeline. Compared to Experiment 3, this integration significantly improved performance across all models: parsing failures dropped below 1.5% for all models, and correction accuracy exceeded 97% in most cases.

GPT-5.4 achieved the strongest performance overall, reaching 98.60% correction accuracy, outperforming all other models and establishing a new state-of-the-art in this setting. Other models also benefited substantially from Kubecurity, with smaller models such as Mistral-7B and Qwen 2.5-7B reaching performance levels comparable to much larger models, highlighting the effectiveness of schema enforcement in compensating for model capacity differences.

The improvement stems from Kubecurity's ability to systematically resolve structural inconsistencies such as indentation errors, missing required fields, and type mismatches by enforcing official Kubernetes JSON schemas. This deterministic validation layer converts near-correct but invalid outputs into fully compliant configurations.

Overall, combining LLM contextual reasoning with schema-based validation proves highly effective, yielding configurations that are syntactically valid, semantically consistent, and fully compliant with Kubernetes standards while substantially reducing the presence of detected security misconfigurations. The inclusion of GPT-5.4 further confirms that stronger foundation models benefit even more from schema-guided correction, achieving near-perfect performance.

## 5 Implications

This study highlights important implications for Kubernetes configuration management. For practitioners, the persistence of misconfigurations demonstrates the limits of manual correction and underscores the need for automated pipelines that combine LLM-based reasoning with schema validation, such as Kubecurity, to enhance reliability and security. Continuous validation and automated remediation should be embedded into deployment workflows.

For tool developers, the findings emphasize the importance of standardizing severity definitions and integrating schema-guided correction mechanisms to enable consistent, interoperable, and self-healing systems.

For researchers, the results call for shared benchmarks, labeled datasets, and systematic evaluations to advance automated correction methods. Hybrid approaches that combine LLM reasoning with formal validation appear particularly promising for strengthening cloud-native system reliability.

Overall, the study advocates a continuous, validation-driven approach to configuration management to reduce human error and improve Kubernetes stability.

## 6 Threats to Validity

The evaluation was conducted on a finite dataset of Kubernetes configurations, detection tools, and Large Language Models (LLMs), which may limit the generalizability of the results. Nevertheless, the dataset was constructed from large-scale real-world Stack Overflow discussions and publicly available GitHub repositories using a structured multi-phase collection and validation process. While the study mainly focuses on Helm chart repositories and may not cover the entire spectrum of Kubernetes configurations, these repositories are widely used in both practice and prior research. In addition, taxonomy construction, rule mapping, and cluster labeling involved manual validation and consensus between two practitioners to improve consistency and reduce noise. Finally, this work focuses on static structural and policy-level analysis through schema validation and state-of-the-art detection tools. Therefore, although repaired configurations satisfy static validation constraints, this does not fully guarantee semantic correctness, runtime safety, or behavioral preservation in real Kubernetes deployments. Extending the evaluation toward dynamic runtime validation and deployment-level testing remains an important direction for future work.

## 7 conclusion

This research paper introduces a comprehensive taxonomy of Kubernetes misconfigurations and analyzes their evolution across Helm chart development phases, showing that while operational issues often decrease, critical security misconfigurations may persist. The study further evaluates LLM-based automated correction and proposes Kubecurity, a schema-guided validation framework that reinforces structural and semantic accuracy. The hybrid approach achieves high correction performance and full Kubernetes compliance, significantly reducing persistent and newly introduced misconfigurations. Overall, the findings provide an empirical foundation for understanding configuration reliability in cloud-native systems and support future research on automated correction and resilience mechanisms in Kubernetes environments.